\documentclass[reprint, superscriptaddress, twocolumn, amsmath, amssymb, prl]{revtex4-1}

\usepackage{graphicx}
\usepackage{dcolumn} 
\usepackage{bm}
\usepackage[colorlinks=true, citecolor=magenta, linkcolor=magenta]{hyperref}

\usepackage{mathrsfs}
\usepackage{color}
\usepackage{soul}
\usepackage{color}

\definecolor{orange}{rgb}{1.0, 0.5, 0.0}
\definecolor{violet}{rgb}{0.78,0.08, 0.52}
\definecolor{green}{rgb}{0.11, 0.35, 0.02}
\definecolor{bluebell}{rgb}{0.64, 0.64, 0.82}
\definecolor{capri}{rgb}{0.0, 0.45, 0.73}

\usepackage{braket}

\usepackage{lipsum}
\usepackage{xargs}
\usepackage{setspace}

\usepackage{float}
\usepackage{booktabs}

\newcommand{\ketbra}[2]{\ket{#1}\!\bra{#2}}

\newcommand{\target}{\text{tgt}}

\newcommand{\Avg}[1]{\ensuremath{\langle #1 \rangle}}

\begin{document}

\title{Dynamically Enhanced Two-Photon Spectroscopy}

\def\ARL{DEVCOM Army Research Laboratory, Adelphi, MD 20783}
\def\UCBer{Department of Physics, University of California, Berkeley, CA 94720}

\author{Sebastian C. Carrasco}
  \email{seba.carrasco.m@gmail.com}
  \affiliation{\ARL}

\author{Sean Lourette}
  \affiliation{\ARL}
\affiliation{\UCBer}

\author{Ignacio Sola}
    \email{isolarei@ucm.es}
    \affiliation{Departamento de Qu\'{\i}mica F\'{\i}sica, Universidad Complutense, 28040 Madrid, Spain}

\author{Vladimir S. Malinovsky}
  \affiliation{\ARL}

\date{\today}

\begin{abstract}
    A novel quantum control protocol utilizing two-photon processes with trigonometric pulse modulation is developed, enabling the intermediate state population's dynamic elimination (DE). The proposed DE technique excels at single-photon resonance in contrast to the well-known adiabatic elimination (AE) regime, which requires large single-photon detuning and strong pulses. The DE approach outperforms AE in efficiency and exhibits enhanced resilience to one-photon detuning since the key control parameter, the effective two-photon Rabi frequency, inversely proportional to the modulation frequency, does not depend on the one-photon detuning. The comprehensive analysis of population transfer and Ramsey interferometry demonstrates the protocol's superiority, achieving enhanced signal-to-noise ratios and higher fidelities with respect to existing two-photon methods.
\end{abstract}

\maketitle

Quantum control techniques are crucial for precise manipulation of quantum systems. Much of the success of quantum technologies stems from our ability to manipulate two-photon transitions. The most universal approach involves stimulated Raman transitions (SRT), where two strong fields, far detuned from single-photon transitions, favor the two-photon transition via an intermediate state. This concept has been instrumental in numerous breakthroughs, including laser cooling~\cite{KasevichPRL1992, ReichelPRL1995, DavidsonPRL1994, MonroePRL1995}, quantum clocks~\cite{AldenPRA2014, RouraPRX2020}, Rydberg blockade~\cite{UrbanNP2009}, atom interferometry~\cite{KasevichPRL1991, KasevichPRL1991vel, MolerPRA1992}, coherent population trapping~\cite{ArimondoLNC1976, GrayOL1978}, and spin squeezing~\cite{Schleier-smithPRA2010, LerouxPRL2010}. In the far-detuning limit, an effective two-level system emerges between states connected by a two-photon transition through an intermediate state. This is often called adiabatic elimination (AE).

For coherent population transfer via two-photon transitions, alternative schemes exist, notably stimulated Raman adiabatic passage (STIRAP)~\cite{VitanovRMP2017}. STIRAP transfers population adiabatically through a dark state created by two fields, leaving the intermediate state unoccupied~\cite{GaubatzJCP1990, Sola2018}. This technique has far-reaching applications in atomic and molecular physics~\cite{GaubatzCPL1988, PilletPRA1993}, trapped-ion physics~\cite{SorensenNJP2006}, optics~\cite{LonghiLPR2009}, superconducting circuits~\cite{KumarNC2016}, entanglement generation~\cite{ChangPRL2020}, optomechanics~\cite{FedoseevPRL2021}, and qubit manipulations~\cite{ToyodaPRA2013, BeterovPRA2013, SetiawanPRA2023}. Its versatility and efficacy make STIRAP a valuable tool in quantum technology.

Despite the remarkable progress in two-photon transition techniques, existing methods are scarce. This work bridges this gap by introducing a novel technique for achieving flexible, high-fidelity two-photon transitions. Our proposed mechanism employs two overlapping pulses with oscillating envelopes in a relative phase offset. This design exploits zero-area pulses~\cite{VasilevPRA2006, RangelovPRA2012} to suppress single-photon transitions, while the phase difference selectively enhances two-photon transitions. By eliminating intermediate state population, our scheme boosts the overall fidelity of the transition. Notably, its analytical solution provides a complete description of the dynamics, guaranteeing the intermediate state's vacancy at completion.

When the modulation frequency $\omega_e$ exceeds $1/t_p$ ($t_p$ being the pulse duration), the addressed three-level system effectively reduces to a two-level system, mirroring AE behavior, with the intermediate state remaining unoccupied during transitions, but generating a dynamical elimination (DE). This similarity prompts a comparative analysis of both techniques. Notably, we find distinct effective Rabi frequencies: $\Omega_\text{eff}(t) = \Omega^2(t) / (4\omega_e)$ for our DE approach, versus $\Omega_\text{eff}^{\text{AE}}(t) = \Omega^2(t) / (2\Delta)$ for AE. This reveals modulation playing a role analogous to single-photon detuning $\Delta$ in our scheme.

Our technique exhibits enhanced robustness against variations in the single-photon detuning, $\Delta$, as its effective Rabi frequency is independent of $\Delta$. Numerical analysis confirms this intuition, showing population transfer fidelities up to 35 times more insensitive to single-photon detuning variations in comparison to AE. Inspired by recent Ramsey interferometry experiments employing STIRAP~\cite{Lourette2024}, we explore what happens when using DE or AE for similar purposes. Both prove excellent candidates, surpassing STIRAP in complementary regimes.

Our approach operates without requiring single-photon detuning, although it remains effective in its presence, except when $\Delta$ approaches $\omega_e$. This positions our technique as complementary to AE, operating in distinct regimes. It also offers advantages when small pulse areas are desired, achieving complete population transfer within this limit. We anticipate our technique sharing applications with AE and STIRAP, given their common properties. Furthermore, it may enable new techniques, such as combining it with STIRAP to cancel non-adiabatic terms~\cite{GongPRA2024} or with rapid adiabatic passage~\cite{SolaPRA1999, malinovsky2003momentum, UnanyanPRL2001, PeikPRA1997,KuznetsovaPhysScr2014,MalinovskayaOptLett2017, CarrascoPRL2024} to create an adiabatic population transfer method leveraging modulation-induced coupling.

\begin{figure}
    \centering
    \includegraphics[scale=0.85]{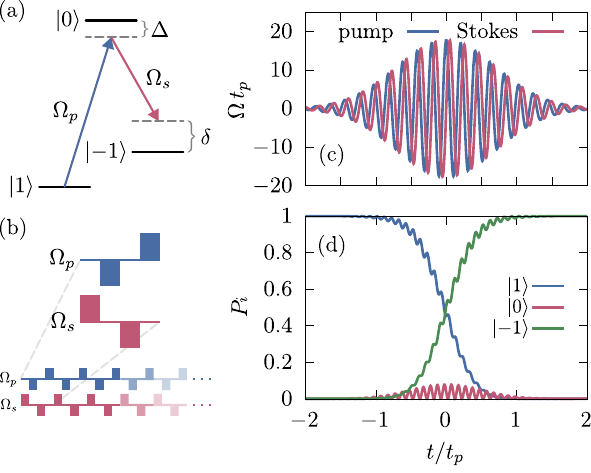}
    \caption{Pulses and population dynamics. (a) shows the schematics of a three-level system, the pump and Stokes pulses, the single-photon detuning $\Delta$ and the two-photon detuning $\delta$. (b) shows a Stokes-pump-Stokes-pump sequence that can drive population between $\ket{\pm 1}$. (c) shows Gaussian pump and Stokes pulses modulated with a relative phase of $\pi/2$. The envelope area is $10\pi$ and the effective pulse area is $\pi$. (d) shows the population dynamics that the pulses in (c) generate.}
    \label{fig:dyn}
\end{figure}

{\it Theoretical model.---}We consider a three-level system composed by two ground states $\ket{\pm 1}$ and an intermediate state $\ket{0}$ (see Fig.~\ref{fig:dyn}(a)). In the field-interaction representation under the rotating wave approximation (RWA) the Hamiltonian of the system driven by two fields has the form
\begin{equation} \label{Eq:Ham}
    H(t) = \frac{1}{2}
    \begin{pmatrix}
        0 & \Omega_p(t) & 0 \\
        \Omega_p(t) & 2\Delta & \Omega_s(t) \\
        0 & \Omega_s(t) & - 2 \delta
    \end{pmatrix} \,,
\end{equation}
where $\Delta$ is the single-photon detuning, $\delta$ the two-photon detuning, and $\Omega_p(t)$ and $\Omega_s(t)$ are the Rabi frequencies of the pump and Stokes fields.

Illustrating our approach, we consider the resonant case ($\delta = \Delta = 0$) with a sequence of four weak, independent pulses: Stokes-pump-Stokes-pump, where one of the pump and one Stokes pulses have a $\pi$ phase shift added to its Rabi frequency (see Fig.~\ref{fig:dyn}(b)). All pulses have equal area $S$. The time-evolution operator for this sequence is
\begin{equation} \label{Eq:Unitary}
    U \approx \left(I - i H_s S\right)\left(I + i H_p S \right)\left(I + i H_s S \right)\left(I - i H_p S \right) \,,
\end{equation}
where $H_p = \frac{1}{2}\ket{1}\!\bra{0} + \text{H.c.}$ and $H_s = \frac{1}{2} \ket{-1}\!\bra{0} + \text{H.c.}$ are the pump and Stokes pulse Hamiltonians.
Simplifying to first order yields
$$U = I - \frac{i}{2} H_t S^2 \,, $$ 
where $H_t=2i[H_s, H_p]=\frac{i}{2}\ket{-1}\!\bra{1} + \text{H.c.}$. This effective target Hamiltonian ($H_t$) enables direct coupling between the system ground states $\ket{\pm 1}$, facilitating complete control within this subspace. Repeating this sequence (or generating a pulse train~\cite{MalinovskayaJOSAB2013, MalinovskayaCPL2016, LiuJMO2018, SolaJPB2022}) avoids ancillary state population. Moreover, the net zero pulse area ---resulting from balanced positive and negative pump and Stokes pulses--- eliminates intermediate state excitation.

We extend our concept by introducing time-dependent Rabi frequencies: $\Omega_p(t) = \Omega(t) \sin(\omega_e t)$ and $\Omega_s(t) = \Omega(t) \cos(\omega_e t)$, where $\omega$ is the modulation frequency, defining the dynamical elimination two-photon control technique. Utilizing average Hamiltonian theory (AHT)~\cite{HaeberlenPR1968, berman2011principles, BrinkmannCMR2016, oon2024beyond}, we derive an effective Hamiltonian for this scenario.

The effective Hamiltonian for a single period $T = 2\pi/\omega_e$ is approximated as $H_\text{eff} \approx H^{(1)} + H^{(2)}$, where $H^{(1)}$ and $H^{(2)}$ represent the first and second orders of the Magnus expansion~\cite{MagnusCPAM1954}. These terms sufficiently describe the system's dynamics. The first-order term, 
\begin{equation}
    H^{(1)} = \frac{1}{T} \int\limits_0^T H(t_1) dt_1 = \Delta \ket{0}\!\bra{0} + \delta \ket{-1}\!\bra{-1}  \, ,
\end{equation}
yields phase accumulation governed by detunings, as if $\Omega (t)$ were zero. Pulse area cancellation eliminates state coupling.
The second-order term, 
\begin{align}
    H^{(2)} &= \frac{1}{2i T} \int\limits_0^T dt_1\int\limits_0^{t_1} dt_2 [H(t_1), H(t_2)] \nonumber \\
    &= \frac{\Omega^2(t)}{4 \omega_e} H_t + \frac{\Delta \Omega(t)}{2\omega_e}  (\ket{1}\!\bra{0} +  \ket{0}\!\bra{1}) \, ,
\end{align}
reveals non-zero coupling between previously uncoupled states $\ket{\pm 1}$. Under resonant conditions, the effective Hamiltonian (up to second order) simplifies to
\begin{equation}
    H_\text{eff} = \frac{\Omega^2(t)}{8 \omega_e} \left[i\ket{-1}\!\bra{1} - i\ket{1}\!\bra{-1}\right] = \frac{\Omega^2(t)}{4 \omega_e} H_t \, .
\end{equation}
This expression mirrors our previous Taylor expansion and pulse sequence results, demonstrating zero-area pulses decouple the intermediate state while ground state coupling arises from pump and Stokes Hamiltonian commutators.

The effective Rabi frequency, $\Omega_\text{eff}(t) = \Omega^2(t)/(4 \omega_e)$, exhibits quadratic dependence on pump and Stokes Rabi frequencies, analogous to the SRT (or AE) scheme but with $\omega_e$ replacing $\Delta$. Notably, $\omega_e$ is a pulse parameter, facilitating precise control. Assuming a Gaussian envelope, $\Omega(t)=\Omega_0 e^{-t^2/t_p^2}$, yields an effective Rabi frequency leading to a quadratic effective area rule: $S_\text{eff} = \sqrt{2} S^2 / (8 \sqrt{\pi} \, \omega_e t_p)$ where $S = \int_{-\infty}^\infty \Omega(t) dt = \sqrt{\pi}  \Omega_0 \, t_p$,  is the envelope area. Specifying pulse area and effective pulse area (or modulation frequency $\omega_e$) fully characterizes the pulses.

The effective interaction is robust under small variations in the pulse parameters. For instance, adding a phase $\phi$ to one of the modulating functions, {\it e.g.} $\cos(\omega_e t +\phi)$ induces a factor $\cos\phi$ to $\Omega_\text{eff}(t)$, as $\Omega_\text{eff}(t) \rightarrow \cos(\phi)\Omega_\text{eff}(t)$. While a $\pi/2$ phase between the trigonometric functions maximizes $\Omega_\text{eff}(t)$, phase drifts can be compensated increasing the area $S$ as long as there is any shift between the modulating functions. Additionally, relative variations in the pulse amplitudes parameterized as $\Omega_p(t) = \cos(\alpha) \Omega(t)$,  $\Omega_s(t) = \sin(\alpha) \Omega(t)$, also induce a factor $\sin(2\alpha)$ in the effective coupling, $\Omega_\text{eff} \rightarrow \cos(2\alpha)\Omega_\text{eff}(t)$ (see Supplementary material).

Eq. \eqref{Eq:Ham} also has an analytical solution for resonant conditions in the adiabatic limit $\omega_e |\dot{\Omega}(t)| /\Omega^2_e(t) \ll \Omega_e(t)$ where $\Omega_e(t)=\sqrt{4\omega_e^2+ \Omega^2(t)}$. If the initial state is $\ket{\Psi(0)}=\ket{1}$, the wave function at time $t$ is given by
\begin{equation} \label{eq:analytical}
    \ket{\Psi(t)} =
    \begin{pmatrix}
        \frac{2 \omega_e}{\Omega_e(t)}  \cos \frac{\Lambda(t)}{2} \cos \omega_e t 
+ \sin \frac{\Lambda(t)}{2} \sin \omega_e t  \\ 
i \frac{ \Omega (t)}{\Omega_e(t)} \cos \frac{\Lambda(t)}{2} \\ 
\sin \frac{\Lambda(t)}{2}  \cos \omega_e t - \frac{2 \omega_e}{\Omega_e(t)}  \cos \frac{\Lambda(t)}{2} \sin \omega_e t 
    \end{pmatrix}
\end{equation}
with $\Lambda(t) = \int_0^t dt'\Omega_e(t')$. We include the derivation in the supplementary material. Notably, Eq. \eqref{eq:analytical} guarantees the population in the intermediate state is zero at the end of the pulses as long as the $\Omega(t)$ goes to zero at the end and $\Omega(t)$ is smooth enough to satisfy adiabatic conditions.  In that case,
\begin{equation} \label{eq:analytical2}
    \ket{\Psi(t)} =
    \begin{pmatrix}
        \cos \left( \frac{\Lambda(t)}{2} - \omega_e t\right),
        0,
        \sin \left( \frac{\Lambda(t)}{2} - \omega_e t\right) 
    \end{pmatrix}^t \ ,
\end{equation}
which corresponds to Rabi oscillations between $\ket{\pm 1}$. Besides, Eq. \eqref{eq:analytical2} gives an exact expression for the effective Rabi frequency $\Omega_\text{eff}(t) = \sqrt{4 \omega_e^2 + \Omega^2(t)} - 2 \omega_e$, which in the limit $\omega_e \gg \Omega(t)$ becomes the Magnus expansion result $\Omega_\text{eff}(t) = \Omega^2(t)/(4\omega_e)$. Back to Eq. \eqref{eq:analytical}, it also demonstrates that the intermediate population of the state $\ket{0}$ goes to zero if $\Omega (t) \gg \Omega_e (t)$, which occurs in the limit $\omega_e \gg \Omega(t)$, thus exhibiting consistency with the Magnus expansion result.

{\it Numerical results.---}Figure~\ref{fig:dyn}(c) illustrates an example of the pump and Stokes pulses we propose. These pulses have an envelope area of $S=10\pi$ and an effective area of $S_\text{eff}=\pi$. Figure \ref{fig:dyn}(d) demonstrates population transfer between the ground states. To highlight the periodic nature of the effective Hamiltonian, we intentionally selected a slow modulation frequency. As a result, some intermediate state excitation occurs. However, this excitation rapidly diminishes as the area and modulation frequency increase, since the intermediate state lacks sufficient time to become excited.

\begin{figure}
    \centering
    \includegraphics[scale=0.85]{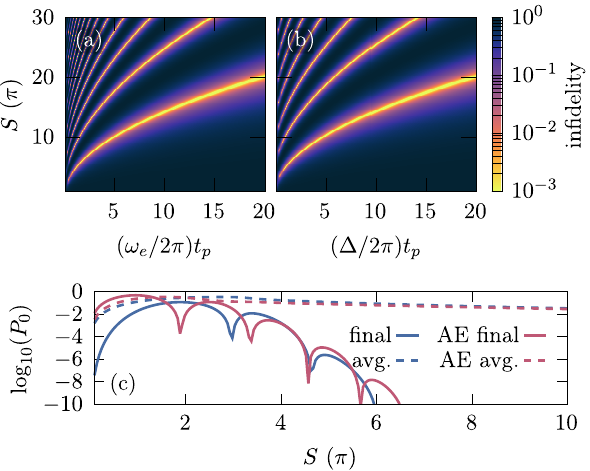}
    \caption{Infidelity and intermediate state population. (a) shows the infidelity of a $\pi$ pulse using the DE technique as a function of the envelope area $S$ and the modulation frequency $\omega_e$. (b) shows the infidelity of a $\pi$ pulse using the AE method as a function of the envelope area $S$ and single-photon detuning $\Delta$. (c) shows a comparison of the average intermediate state population between $t=-t_p$ and $t_p$, and the final intermediate state population for the two techniques.}
    \label{fig:infidelity}
\end{figure}

To provide insight into our technique's behavior, we numerically calculated the infidelity of a $\pi$ pulse, $1-|\!\braket{-1|\Psi(T)}\!|^2$ with $\Psi(T)$ the wavefunction at final time, as a function of envelope area, $S = \int dt' \sqrt{\Omega_p^2(t') + \Omega_s^2(t')}$, and modulation frequency $\omega_e$ (Fig.~\ref{fig:infidelity}(a)). The fidelity exhibits square-root-shaped maxima, confirming the theoretical prediction that $\omega_e$ is proportional to $S^2$. The peak positions align with our theoretical expectations. Moving from bottom to top, each square-root-shaped region corresponds to effective areas of $\pi$, $3\pi$, $5\pi$, etc. Notably, the technique becomes increasingly robust as $S$ increases, evident from the broadening peaks. The right panel shows a similar calculation but using stimulated Raman transitions but varying $\Delta$ instead of $\omega_e$ (which doesn't exist in the case of AE). The plot looks almost identical due to the similarity of the effective Rabi frequency for this case, $\Omega_\text{eff}^{\text{AE}}(t) = \Omega^2(t) / (2\Delta)$. The bottom panel of Fig.~\ref{fig:infidelity} shows a comparison between the average intermediate-state population, $\Avg{P_0} = \int_{-t_p/2}^{t_p/2} \!dt' \left| \langle 0 | \psi(t') \rangle \right|^2 / t_p$, and final population, for both techniques. We observe that these techniques follow a similar overall behavior, $\Avg{P_0}$ peaks around for low pulse areas and then decreases. The final intermediate-state population initially peaks but rapidly drops to values below $10^{-5}$. These results demonstrate that our scheme performs optimally with increasing envelope area and modulation frequency, effectively minimizing intermediate state excitation.

\begin{figure}
    \centering
    \includegraphics[scale=0.85]{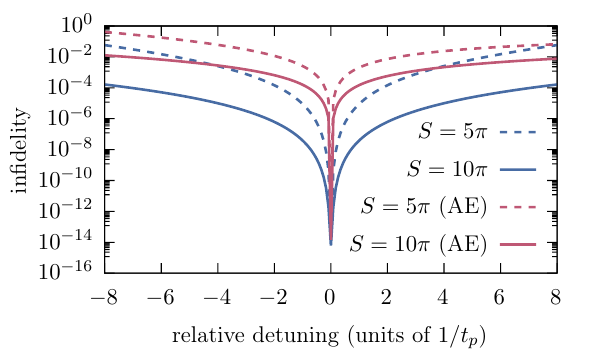}
    \caption{Infidelity comparison for $\pi/2$ pulses created using DE and AE pulses. When simulating DE, we set $\omega_e$ to minimize the infidelity for each pulse area considered here ($S = 5\pi$ and $S=10\pi$) using a reference value of single-photon detuning $\Delta=0$. When simulating AE, we set the reference value of $\Delta$ that minimize the infidelity. Then, we add an additional single-photon detuning to make the comparison. Thus, the simulations employ the reference value of $\Delta$ when the relative detuning is equal to zero. We define the infidelity as $1-|\!\braket{\Psi_\target|\Psi(T)}\!|^2$, where $T$ is the final time and $\ket{\Psi_\target} = (\ket{1} + \ket{-1})/\sqrt{2}$.}
    \label{fig:comparison}
\end{figure}

{\it Non-resonant scenario.---}When single-photon ($\Delta$) and two-photon detunings are non-zero, AHT predicts deviations from resonant dynamics at first and second order of the Magnus expansion. Notably, our technique differs from AE excitation in two key aspects: the effective Rabi frequency expression replaces $\Delta$ with the modulation frequency ($\omega_e$) in DE, and pulse areas for one-photon transitions are zero. These factors enhance the dynamical elimination of intermediate state population, outperforming adiabatic elimination for similar pulse areas (Fig.~\ref{fig:comparison}). Consequently, DE exhibits increased robustness against single-photon detuning variations. Figure~\ref{fig:comparison} compares DE and AE using two pulse areas. DE calculations demonstrates: slower infidelity decay with small detuning deviations and similar maximal fidelity due to depleted $\ket{0}$ state amplitude. To give a quantitative idea of the robustness to detuning variations, when the pulse area is $5 \pi$, the dynamic elimination technique retains infidelities below $10^{-4}$ for $3.9$ times larger relative detunings than AE. If the pulse area is $10 \pi$, that value increases to $35.5$. This leads to improved average fidelity for gates operating under single-photon detuning distributions (e.g., Doppler broadening). DE surpasses AE in non-resonant two-photon absorption for similar pulse areas. This advantage extends to spectroscopic techniques requiring precise two-level system dynamics isolation, such as Ramsey spectroscopy. 

{\it Applications to three-level Ramsey interferometry.---}Ramsey interferometry typically involves preparing a probe state that maximizes sensitivity to the estimated parameter while minimizing unwanted sensitivity. This is followed by free evolution, encoding phases, and decoding with unitary operations, culminating in measurement.
In three-level systems, the rotating frame's free evolution operator is
\begin{equation} \label{Eq:Ufree}
    U_\text{free} = 
    \ketbra{1}{1} + e^{i\delta\tau} \ketbra{-1}{-1} + e^{-i\Delta \tau} \ketbra{0}{0} \,, 
\end{equation}
where $\tau$ is the free evolution time. A relevant scheme is the Double-Quantum (DQ) Ramsey scheme~\cite{BarryRMP2020}, employing an equal superposition of $\ket{\pm 1}$ as the probe state. This makes the scheme sensitive to two-photon detuning $\delta$ and insensitive to single-photon detuning $\Delta$. However, creating this superposition without exciting $\ket{0}$ state poses a significant challenge, and any excitation would introduce detrimental sensitivity to $\Delta$. Existing implementations, such as Ref.~\cite{JarmolaSA2021}, utilize standard Rabi flopping pulses for state manipulation in nuclear-spin transitions of $^{14}$N within nitrogen-vacancy (NV) color centers in diamond. These approaches require canceling the unwanted sensitivities due to imperfect probe states caused imperfections in the control pulses (RF field  gradients in \cite{Lourette2024}). A follow-up work, Ref.~\cite{Lourette2024}, proposes using STIRAP to transfer population between $\ket{\pm 1}$ while avoiding $\ket{0}$ excitation. However, large pulse areas limit this technique's effectiveness in suppressing $\ket{0}$ population, only partially diminishing unwanted sensitivities. Our DE technique offers an alternative solution, decoupling $\ket{0}$ from the dynamics and preventing its excitation.

\begin{figure}
    \centering
    \includegraphics[scale=0.85]{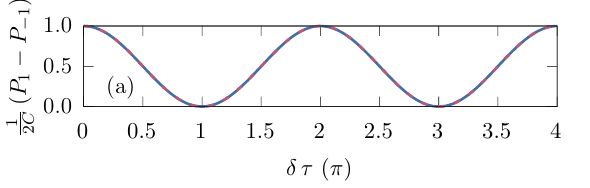}
    \includegraphics[scale=0.85]{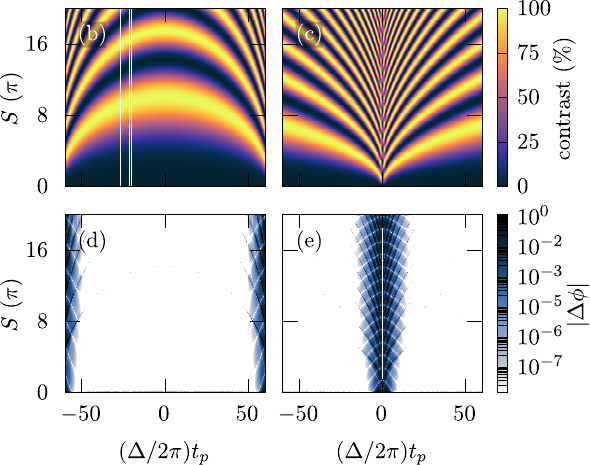}
    \caption{Ramsey interferometry using DE and AE pulses. (a) shows two Ramsey fringes normalized by the contrast. The solid line represent ideal conditions for the DE $\pi/2$-pulses ($S = 10\pi$ and $\Delta \, t_p = 0$) while the dashed line deviates from the ideal conditions ($S = 5\pi$ and $(\Delta/2\pi) t_p = 20$). Nevertheless, both curves coincide except for the amplitude and there isn't any appreciable phase shift. (b) shows the contrast $C$, defined as the amplitude of the fringes, as a function of the same parameters. (c) repeats the calculation of panel (b) but using AE pulses. (d) shows the phase shift $\Delta \phi$ of the Ramsey fringes as a function of envelope area $S$ and normalized detuning $\Delta \, t_p$. (e) repeats the calculation of panel (d) but using AE pulses. All DE pulses in the figure use a modulation frequency $(\omega_e/2\pi) t_p = 10$.
    }
    \label{fig:ramsey}
\end{figure}

Figure~\ref{fig:ramsey} presents numerical simulations of Ramsey interferometry using DE pulses, employing experimentally relevant parameters from~\cite{JarmolaSA2021, Lourette2024}. Fig.~\ref{fig:ramsey}(a) displays normalized Ramsey fringes for population differences between $\ket{\pm 1}$ states. The solid line represents ideal parameters creating the desired probe state, while the dashed line corresponds to non-optimal pulse areas (half of the optimal value) and off-resonant conditions ($(\Delta/2\pi) t_p = 20$). Despite differences, normalization reveals an identical fringe shape without relative phases, which shows the robustness that DE brings to the Ramsey fringes. Fig.~\ref{fig:ramsey}(b) plots the computed Ramsey fringes contrast, highlighting extensive regions of maximum contrast in the parameter space. This demonstrates the robustness conferred by DE. Indeed, when comparing to the AE technique (Fig.~\ref{fig:ramsey}(c)), we observe a clear dependence on $\Delta$; the optimal value changes with $\Delta$ linearly, while for DE, the correction is quadratic. Fig.~\ref{fig:ramsey}(d) illustrates phase shifts $\Delta \phi$ in Ramsey fringes as a function of pulse area and single-photon detuning $\Delta$. DE in DQ Ramsey renders fringes insensitive to pulse area and $\Delta$ variations, provided $\Delta$ is not excessively large. Notably, for $t_p = 500$ $\mu$s (Ref.~\cite{Lourette2024}), unwanted phase shifts occur at $\Delta \approx 10$ kHz, surpassing STIRAP's performance. A similar behavior is shown in Fig.~\ref{fig:ramsey}(e) for the AE technique but for a complementary region of parameter space, namely, when $\Delta$ is large.

{\it Summary.---} 
We introduce a novel quantum control technique in a three-level Lambda system utilizing pulses modulated by phase-shifted oscillatory functions. The proposed control protocol provides almost ideal two-photon excitation while suppressing single-photon excitations during population transfer and Ramsey interferometric measurements. The obtained analytical solution for single-photon resonance and an approximate solution for the non-resonant condition help us extract the mechanism and prominent features of the control process. The analytical solution reveals that the intermediate state is always empty at the final time. Additionally, the approximate solution shows that oscillatory modulation suppresses single-photon excitations by making the pump pulse area zero and exponentially reducing the Stokes pulse area with increasing modulation rate.

We demonstrate that sufficiently high modulation frequencies enable the dynamical elimination of the intermediate state. This control process is analogous to adiabatic elimination in stimulated Raman transition schemes but works well in single-photon resonance, whereas adiabatic elimination requires large single-photon detuning and pulse areas. The complementarity of these strategies is evident in the population transfer's dependence on the effective two-photon Rabi frequency, scaling as the squared single-photon Rabi frequency over single-photon detuning for adiabatic elimination and modulation frequency for dynamical elimination.

A key feature of our protocol is the independence of the effective two-photon Rabi frequency (the primary control parameter) from single-photon detuning, which is crucial for DQ Ramsey spectroscopic measurements~\cite{LourettePRA2023,Lourette2024}. Our calculations reveal amplified robustness of Ramsey-fringes contrast to pulse area and single-photon detuning variations, with practically suppressed phase shifts. This enhances sensitivity and robustness in quantum sensors like clocks, gyroscopes, and magnetometers.

The proposed two-photon spectroscopy technique propels advancements in quantum optics, laser spectroscopy, and quantum information research. We anticipate its widespread adoption in atomic, molecular, and spin systems, fostering robust control methods for quantum technology applications, including high-fidelity quantum gates~\cite{MalinovskyPRL2004}, Ramsey spectroscopy, atom interferometry and quantum sensing.

\begin{acknowledgments} 
This research was supported by DEVCOM Army Research Laboratory under Cooperative Agreement Numbers W911NF-21-2-0044 (SCC), W911NF-24-2-0050 (SL) and the Ministerio de Ciencia e Innovación of Spain (MICINN), Grant No. PID2021-122796NB-I00.
\end{acknowledgments} 

\bibliographystyle{apsrev4-2.bst}
\bibliography{refs}

\onecolumngrid

\section*{SUPPLEMENTARY MATERIAL}
\renewcommand{\theequation}{S\arabic{equation}}
\renewcommand{\thefigure}{S\arabic{figure}}
\setcounter{figure}{0}
\setcounter{equation}{0}

\subsection{Different pump and Stokes amplitudes}
For the resonant case (single-photon and two-photon detuning are zero), if we consider 
pump and Stokes amplitudes parameterized as 
$\Omega_p(t) = \cos(\alpha) \Omega(t) \sin(\omega_e t)$ and $\Omega_s(t) = \sin(\alpha) \Omega(t) \cos(\omega_e t)$, we obtain
\begin{equation}
    H_\text{eff} \approx - i \frac{\omega_e}{2 \pi} \int\limits_{-\pi/\omega_e}^{\pi/\omega_e} H(t_1) dt_1 - i \frac{\omega_e}{4 \pi} \int\limits_{-\pi/\omega_e}^{\pi/\omega_e} dt_1 \int\limits_{-\pi/\omega_e}^{t_1} dt_2 [H(t_1), H(t_2)]  = \frac{\sin(2 \alpha) \Omega^2(t)}{4 \omega_e} H_t
\end{equation}
when computing the Magnus expansion up to second order.

\subsection{General relative phase between pump and Stokes}
In resonance, if there is a arbitrary relative phase $\phi$ betweem pump and Stokes, $\Omega_p(t) = \Omega(t) \sin(\omega_e t)$ and $\Omega_s(t) = \Omega(t) \cos(\omega_e t + \phi)$, we obtain 
\begin{equation}
    H_\text{eff} \approx - i \frac{\omega_e}{2 \pi} \int\limits_{-\pi/\omega_e}^{\pi/\omega_e} H(t_1) dt_1 - i \frac{\omega_e}{4 \pi} \int\limits_{-\pi/\omega_e}^{\pi/\omega_e} dt_1 \int\limits_{-\pi/\omega_e}^{t_1} dt_2 [H(t_1), H(t_2)]  = \frac{\cos(\phi) \Omega^2(t)}{4 \omega_e} H_t
\end{equation}
when calculating the Magnus expansion up to second order.

\subsection{Next order correction}
For the resonant scenario, the third-order term is given by
\begin{equation}\label{Mag3}
    H^{(3)} = -i \frac{\omega_e}{12 \pi} \int\limits_{-\pi/\omega_e}^{\pi/\omega_e} dt_1 \int\limits_{-\pi/\omega_e}^{t_1} dt_2 \int\limits_{-\pi/\omega_e}^{t_2} dt_3 \left([H(t_1), [H(t_2), H(t_3)]] + [H(t_3), [H(t_2), H(t_1)]] \right) = \frac{\pi}{36 \omega_e^2} H_t \, .
\end{equation}
Note that the Hamiltonian in Eq.~(\ref{Mag3}) directly couples  the $\ket{\pm 1}$ states.

\subsection{Adiabatic analytical result in the rotating frame}

For the resonant case, the Hamiltonian in Eq.~(\ref{Eq:Ham}) takes the form
\begin{equation}\label{RWA_1}
H (t) =
    \frac{1}{2} 
    \begin{pmatrix}
       0 & \Omega (t)  \sin \omega_e t & 0 \\
    \Omega (t) \sin \omega_e t & 0 & \Omega (t) \cos \omega_e t \\
    0 & \Omega (t) \cos \omega_e t & 0  
    \end{pmatrix}\,.
\end{equation}
Making the transformation
\begin{equation}\label{Trans}
R (t)  = 
\begin{pmatrix}
\cos \omega_e t & 0 &  \sin \omega_e t \\
0 & 1 & 0 \\
-\sin \omega_e t & 0 & \cos \omega_e t 
\end{pmatrix} \,,
\end{equation}
we obtain
\begin{align}\label{RWA_2}
\tilde{H} (t) &=R^{-1}(t) H (t) R(t) - i R^{-1}(t) \dot{R}(t) \nonumber \\ 
&= \frac{1}{2} 
\begin{pmatrix}
 0 & 0 & -2i \omega_e\\
0 & 0 & \Omega (t) \\
 2i \omega_e& \Omega (t) & 0 
\end{pmatrix}
\,.
\end{align}

Using the unitary transformation
\begin{equation}\label{Trans2}
U(t) = 
\begin{pmatrix}
    - \frac{\sqrt{2} \, i \omega_e}{\Omega_e(t)} & \frac{i \Omega (t)}{\Omega_e(t)}  & - \frac{\sqrt{2} \, i \omega_e}{\Omega_e(t)}  \\
    \frac{ \Omega (t)}{\sqrt{2} \Omega_e(t)}   & \frac{2 \omega_e}{\Omega_e(t)}  & \frac{ \Omega (t)}{\sqrt{2} \Omega_e(t)} \\
    -\frac{ 1}{\sqrt{2} }  & 0 & \frac{ 1}{\sqrt{2} } 
\end{pmatrix}
\end{equation}
to diagonalize the Hamiltonian in Eq.~\eqref{RWA_2}, we get
\begin{align}\label{DiagH}
   \bar{H}(t) &= D(t) - i U^{-1}(t) \dot{U} (t) \nonumber \\
    &= \frac{\Omega_e (t)}{2}
    \begin{pmatrix}
        - 1 & 0 & 0\\
        0 & 0 & 0 \\
        0 & 0 & 1  
    \end{pmatrix} + \sqrt{2} \, i\frac{ \omega_e \dot \Omega(t)}{\Omega_e^2(t)}
    \begin{pmatrix}
        0 & 1 & 0\\
        -1 & 0 & -1 \\
        0 & 1 & 0  
    \end{pmatrix}
\,,
\end{align}
where $D(t) = U^{-1}(t) \tilde{H} (t) U (t)$ is a diagonal matrix, and $\Omega_e(t)=\sqrt{4\omega_e^2+ \Omega^2(t)}$. The second term in Eq.~\eqref{DiagH} represents the non-adiabatic coupling between eigenstates of the Hamiltonian $\tilde{H}$, Eq.~\eqref{RWA_2}, with energies $\pm \Omega_e(t)/2$, and $0$.

Neglecting the non-adiabatic coupling, under the condition $\omega_e |\dot{\Omega} (t)|/\Omega_e^2(t) \ll \Omega_e(t)$, we obtain the following time-evolution operator
\begin{align}\label{EvolFin}
    U_\text{DE}(t) &= R(t) \, U(t) \, \exp\left(-i \int_0^t dt' D(t')\right) \, U^{-1}(0) \, R^{-1}(0) \nonumber \\
     &= \begin{pmatrix}
\frac{2 \omega_e}{\Omega_e(t)}  \cos \frac{\Lambda(t)}{2} \cos \omega_e t 
+ \sin \frac{\Lambda(t)}{2}  \sin \omega_e t & 
i \frac{ \Omega (t)}{\Omega_e(t)}  \cos \omega_e t & 
\cos \frac{\Lambda(t)}{2} \sin \omega_e t  - \frac{2 \omega_e}{\Omega_e(t)}  \sin \frac{\Lambda(t)}{2} \cos \omega_e t \\
i \frac{ \Omega (t)}{\Omega_e(t)} \cos \frac{\Lambda(t)}{2}  & \frac{2 \omega_e}{\Omega_e(t)}  & 
-i \frac{ \Omega (t)}{\Omega_e(t)} \sin \frac{\Lambda(t)}{2}  \\
\sin \frac{\Lambda(t)}{2} \cos \omega_e t - \frac{2 \omega_e}{\Omega_e(t)}  \cos \frac{\Lambda(t)}{2}  \sin \omega_e t  & 
-i \frac{ \Omega (t)}{\Omega_e(t)}  \sin \omega_e t & 
\frac{2 \omega_e}{\Omega_e(t)}  \sin \frac{\Lambda(t)}{2}\sin \omega_e t 
+ \cos \frac{\Lambda(t)}{2} \cos \omega_e t 
\end{pmatrix} \, ,
\end{align}
where $\Lambda(t) = \int_0^t dt'\Omega_e(t')$.

The first column in Eq.~\eqref{EvolFin} represents the time-evolution of the initial state $\ket{1}$, equivalent to Eq.~\eqref{eq:analytical}. 
\end{document}